\documentclass[11pt,a4paper]{article}
\pdfoutput=1
\usepackage{jcappub}

\usepackage{pdflscape}
\usepackage{amsmath}
\usepackage{amssymb}
\usepackage{dcolumn}
\usepackage{bm}
\usepackage{color}
\usepackage{epsfig}
\usepackage{amsfonts}
\usepackage{subfigure}
\usepackage{dcolumn}

\DeclareMathOperator{\tr}{tr}
\def\widebar{\overline}
\def\mathsf{\tr}

\begin{document}

\title{Wormholes minimally violating the null energy condition}

\author[a,b,c,d]{Mariam Bouhmadi-L\'opez}
\author[e,f]{Francisco S. N. Lobo}
\author[e,f,g]{Prado Mart\'{\i}n-Moruno}

\affiliation[a]{Departamento de F\'{i}sica, Universidade da Beira Interior, 6200 Covilh\~a, Portugal}

\affiliation[b]{Centro de Matem\'atica e Aplica\c{c}\~oes da Universidade da Beira Interior (CMA-UBI)}

\affiliation[c]{Department of Theoretical Physics, University of the Basque Country UPV/EHU, P.O. Box 644, 48080 Bilbao, Spain}

\affiliation[d]{IKERBASQUE, Basque Foundation for Science, 48011, Bilbao, Spain}

\affiliation[e]{Centro de Astronomia e Astrof\'{\i}sica da Universidade de Lisboa, Campo Grande, Edif\'{\i}cio C8, 1749-016 Lisboa, Portugal}

\affiliation[f]{Instituto de Astrof\'{\i}sica e Ci\^{e}ncias do Espa\c{c}o, Universidade de Lisboa, OAL, Tapada da
Ajuda, PT1349-018 Lisboa, Portugal.}

\affiliation[g]{Centro Multidisciplinar de Astrof\'{\i}sica - CENTRA, Departamento de F\'{\i}sica, Instituto Superior T\'ecnico, Av. Rovisco Pais 1,1049-001 Lisboa, Portugal}

\emailAdd{mariam.bouhmadi@ehu.es}
\emailAdd{fslobo@fc.ul.pt}
\emailAdd{pmmoruno@fc.ul.pt}

%%%%%%%%%%%%%%%%%%%%%%%%%%%%%%%%%%%%%%%%%%%%%%%%%%%%%%%%
\abstract{
We consider novel wormhole solutions supported by a matter content that minimally violates the null energy condition. 
More specifically, we consider an equation of state in which the sum of the energy density and radial pressure is proportional to  a constant with a value smaller than that of
the inverse area characterising the system, i.e., the area of the wormhole mouth. This approach is motivated by a recently proposed cosmological event, denoted ``the little sibling of the big rip'',  
where the Hubble rate and the scale factor blow up but the cosmic derivative of the Hubble rate does 
not \cite{Mariam}. By using the cut-and-paste approach, we match interior spherically symmetric wormhole solutions to an exterior Schwarzschild geometry, 
and analyse the stability of the thin-shell to linearized spherically symmetric perturbations around static solutions, by choosing suitable properties for the exotic material 
residing on the junction interface radius. Furthermore, we also consider an inhomogeneous generalization of the equation of state considered  above and analyse the respective stability regions. In particular, we obtain a specific wormhole solution with an asymptotic behaviour corresponding to a global monopole.}
%%%%%%%%%%%%%%%%%%%%%%%%%%%%%%%%%%%%%%%%%%%%%%%%%%%%%%%%

\maketitle

\section{Introduction}

The discovery of the current accelerated expansion of the Universe has opened novel research avenues in modern cosmology and has led to the consideration of new scenarios for the future cosmological evolution. 
Even without abandoning general relativity our Universe could be condemned to different kinds of doomsdays  (see \cite{Kamenshchik:2013naa} for a recent review on the topic). From the disaster of the big 
rip \cite{Starobinsky,Caldwell1,Carroll,Caldwell2,BouhmadiLopez:2004me}  or  big freeze \cite{BouhmadiLopez:2006fu,BouhmadiLopez:2007qb,Nojiri:2005sx} singularities to the more benign little 
rip \cite{Stefancic:2004kb,Nojiri:2005sr,BouhmadiLopez:2005gk,Frampton:2011sp} (see also \cite{Ruzmaikina1970} for the first cosmological model with a (past) 
little rip), or little sibling of the big rip recently discussed in \cite{Mariam} it is difficult to continue trusting 
in the peaceful thermal death predicted by most of the pre-acceleration models and the standard $\Lambda$CDM model . 
These doomsdays are commonly related to the consideration of phantom energy as responsible for the accelerated expansion, a dark fluid which violates the null energy condition (NEC) and seems to be favoured by current observational constraints \cite{EscamillaRivera:2011q,Kazin:2014qga,Ade:2013zuv}. It should be pointed out, however, that phantom models could also allow a safe future for our Universe with the thermal death being again the rescuer of the cosmological history \cite{BouhmadiLopez:2004me}. In addition, close to those events quantum effects can be important and can indeed avoid those singularities \cite{Dabrowski:2006dd,Kamenshchik:2007zj,BouhmadiLopez:2009pu,Bouhmadi-Lopez:2013tua}.

The possible existence of phantom energy also leads to the investigation on wormhole physics, since this material content is compatible with these geometries \cite{Sushkov:2005kj,Lobo:2005us}. Far from being an abomination characterizing phantom models, they could be the solution to the doomsday problem. That is, due to the accretion of phantom fluid, they could increase their size faster than the expansion rate of the universe itself, engulfing the whole universe which would travel through the hole in a so-called 
``big trip'' \cite{GonzalezDiaz:2004vv,GonzalezDiaz:2004df,Yurov:2006we}. This big trip would avoid not only the big rip, but any singularity characterized by a divergence of the Hubble parameter \cite{MartinMoruno:2007se} (as it was shown after satisfactorily clarifying \cite{GonzalezDiaz:2007gt}  some criticisms \cite{Faraoni:2007kx} to the treatment leading to this phenomenon). Although as in the black hole case the thermal radiation emitted by wormholes would not be  detectable \cite{MartinMoruno:2009ii,MartinMoruno:2009iu}, one could look for other observational signatures which would be measurable \cite{Cramer, Torres, Torres2, Shatskiy, Pedro, Ana,Harko:2008vy,Harko:2009xf}.

The study of wormhole solutions has been controversial from the beginning \cite{Morris:1988cz,Matt}. 
On one hand, it seems that it would be absurdly easy to ``construct'' a time-machine once one were to find a
wormhole~\cite{Morris:1988tu}. Therefore, one should consider these objects with some caution.
On the other hand, it was thought that exotic matter, needed to maintain their mouths opened, 
is not easy to find in our environment. Nevertheless, one could argue that the phantom energy in cosmological models, responsible for the cosmic speed-up, could provide an inexhaustible source of exotic matter necessary to sustain wormhole geometries. The realization of this fact has led to the study of wormhole solutions
supported by different kinds of phantom fluids (see, for instance \cite{Sushkov:2005kj,Lobo:2005us,Zaslavskii:2005fs,Lobo:2005vc,Lobo:2006ue,Kuhfittig:2006xj,Gonzalez:2008wd,
Gonzalez:2008xk,DeBenedictis:2008qm}).

However, the issue of exotic matter, i.e., matter that violates the NEC, remains a subtle issue \cite{Lobo:2004wq}, and one would prefer to minimize its usage; indeed, solutions minimizing the required amount of this matter have been studied \cite{Kuhfittig:1999nd,Lemos:2003jb,Visser:2003yf,Visser:1989kh,Visser:1989kg}. In fact, in the context of modified gravity it was shown that one may impose that the matter threading the wormhole satisfies the energy conditions, so that it is the effective stress-energy tensor containing higher order curvature derivatives that is responsible for the NEC violation. Thus, the higher order curvature terms, interpreted as a gravitational fluid, sustain these non-standard wormhole geometries, fundamentally different from their counterparts in general relativity  \cite{Lobo:2007qi, Lobo:2008zu, Lobo:2009ip, Garcia:2010xb, MontelongoGarcia:2010xd, Bohmer:2011si,Capozziello:2012hr, Harko:2013yb,Harko:2013aya}. 

In this paper, we consider a new approach most adapted to the current physical conception.  
Although a fluid violating the NEC (and, therefore, all the classical energy conditions) could be driving the dynamics of our  Universe, 
one could think that it is still desirable that such violations not be arbitrarily large.  Therefore, 
we will be interested in minimizing the violation of the NEC by bounding the violation of the inequality associated with this condition, instead of minimizing the use of exotic material.
This notion of not arbitrarily large violations of some energy conditions has been recently explicitly formulated in \cite{Martin-Moruno:2013sfa, Martin-Moruno:2013wfa}. 
It must be noted that when allowing bounded violations of the WEC, one is considering that some observers can measure negative energies 
although not arbitrarily large in absolute value (in \cite{Martin-Moruno:2013sfa, Martin-Moruno:2013wfa} the bounce depends on the characteristics of the system under consideration).
Although a covariant formulation of 
small violations of the NEC similar to that of the WEC formulated in  \cite{Martin-Moruno:2013sfa} is not 
possible, in this paper we will consider a natural formulation by restricting attention to bounded violations which results once our stress energy tensor is 
contracted with the null
vector, that is $p+\rho>-\epsilon^2$.
A fluid of this kind has recently being considered in a cosmological scenario to study the implications which could appear in our universe due to small constant departures from a cosmological constant \cite{Mariam}, that is $\rho+p=-A$ where $A$ is a constant which could be arbitrarily small. It has been shown that this small departure from the cosmological constant case is enough to change the dynamical behaviour of the universe, 
leading to what was named as the ``little sibling of the big rip event'' \cite{Mariam}, i.e., a future event where the Hubble rate and the scale factor blow up but the cosmic derivative of the Hubble rate does not. This abrupt event takes place at an infinite cosmic time where the scalar curvature diverges \cite{Mariam}.

This paper is outlined in the following manner: In section \ref{general}, we summarize some known characteristics of wormholes, emphasizing the use of thin shells to recover a given asymptotic behaviour, rephrasing existing studies \cite{oai:arXiv.org:gr-qc/0506001, Lobo:2005uf, Lobo:2005zu} in a language inspired by references \cite{gravastar, Nadiezhda} in \ref{thinshell}. In section \ref{constantes}, we present the equation of state analysed in reference \cite{Mariam} within a cosmological setup, and consider the possible existence of wormholes supported by this exotic matter. We show that these are not asymptotically flat wormholes solutions, and thus the usage of thin shells is necessary. Next, we study solutions with a constant redshift function in subsection \ref{constantphi} which can be matched to an exterior Schwarzschild geometry, and explore the stability of the system. In section \ref{inhomogeneous}, we introduce an inhomogeneous generalization of the equation of state, considered in the previous section. For these solutions, we assume a particular kind of shape function in subsection \ref{inhshape} which usually leads to asymptotically flat solutions; however, for this case the temporal component of the metric is not well-behaved in this limit and we must to consider again the existence of a thin shell. On the other hand, for a constant redshift function we obtain some solutions that are well-behaved in the asymptotic limit, although they are not asymptotically flat. Finally, in section \ref{summary} we summarize and discuss the results.

%%%%%%%%%%%%%%%%%%%%%%%%%%%%%%%%%%%%%%%%%%%%%%%%%%%%%%%%%%%%%%%%%%%%%%%%%%%%%%%%%%%%%%%%%
%%%%%%%%%%%%%%%%%%%%%%%%%%%%%%%%%%%%%%%%%%%%%%%%%%%%%%%%%%%%%%%%%%%%%%%%%%%%%%%%%%%%%%%%%

\section{General considerations}\label{general}

\subsection{Metric and field equations}

A wormhole is a connection between two universes or a short-cut between two separate regions of one
universe. Thus, it consists of a throat connecting two regions (usually assumed to be asymptotically flat), 
that is~\cite{Morris:1988cz,Matt,Lobo:2007zb}
\begin{equation}\label{metric-l}
 ds^2=-e^{2\Phi(l)}dt^2+{\rm d}l^2+r^2(l)\left(d\theta^2+
 \sin^2d\varphi^2\right),
\end{equation}
where $l$ is the proper radial distance with the range $-\infty<l<\infty$. The function $\Phi(l)$ must be
finite everywhere to avoid the existence of horizons. The radius of the throat is the
minimum of the function $r(l)$, $r_0$. If we suppose that $l(r_0)=0$, then $l<0$ and
$l>0$, respectively, cover the two connected regions. 
If one additionally requires both regions to be asymptotically flat, the functions
$r(l)/|l|$ and $\Phi(l)$  must equal unity and
constant when $l\rightarrow\pm\infty$, respectively. 
Metric~(\ref{metric-l}) can be written using two patches of Schwarzschild coordinates as
\begin{equation}\label{metrica}
ds^2=-e ^{2\Phi(r)} \,dt^2+\frac{dr^2}{1- b(r)/r}+r^2 \,(d\theta^2+\sin ^2{\theta} \, d\phi ^2),
\end{equation}
with $r_0\leq r\leq\infty$, $\Phi(r)$ and $b(r)$ are the redshift function and
shape function, respectively,
and the proper radial distance is defined as
\begin{equation}\label{lr}
 l(r)=\pm\int^r_{r_0}\frac{dr^*}{\sqrt{1-b(r^*)/r^*}}.
\end{equation}
The proper radial distance, $l(r)$, is finite throughout the whole spacetime, thus $b(r)/r<1$ for $r>r_0$,
and $b(r_0)=r_0$ at the throat where the embedded surface is vertical.
Moreover, the embedding surface $z(r)$, given by 
\begin{equation}\label{zr}
 \frac{dz}{dr}=\pm\frac{1}{(r/b-1)^{1/2}},
\end{equation}
must flare outward. That is, it must have a decreasing derivative in the upper half, which implies 
$b'(r)<b(r)/r$ at or near the throat. Thus,
 $b'(r_0)<1$, which can also be obtained by demanding the trapping horizon to 
be outer \cite{Hayward:2009yw,MartinMoruno:2009iu}.

Consider the most general stress-energy tensor compatible with the spacetime~(\ref{metrica}) given by $T^\mu{}_\nu={\rm diag}\left[-\rho(r),\,p_r(r),\,p_t(r),\,p_t(r)\right]$, where $\rho$, $p_r$ and $p_t$ are the energy density, the radial pressure and tangential pressure, respectively. Thus, the Einstein equations
(with $c=G=1$)
imply \cite{Morris:1988cz,Matt,Lobo:2007zb}
\begin{eqnarray}
\rho&=&\frac{1}{8\pi} \, \frac{b'}{r^2},
\label{rho}\\
p_r &=&\frac{1}{8\pi} \, \left[2 \left(1-\frac{b}{r}\right) \frac{\Phi'}{r}- \frac{b}{r^3}\right],
\label{pr}\\
p_t&=&\frac{1}{8\pi} \left(1-\frac{b}{r}\right)\left[\Phi ''+(\Phi')^2- \frac{b'r-b}{2r^2(1-b/r)}\Phi'-\frac{b'r-b}{2r^3(1-b/r)}+\frac{\Phi'}{r} \right],
\label{pt}
\end{eqnarray}
and the conservation equation, $\nabla_\mu T^\mu{}_\nu=0$, leads to
\begin{equation}\label{conservacion}
 p'_r=\frac{2}{r}\left(p_t-p_r\right)-\left(\rho+p_r\right)\Phi',
\end{equation}
which can also be obtained from equations (\ref{rho})--(\ref{pt}).
Thus, we have five unknown functions, $\rho(r),\,p_r(r),\,p_t(r),\,b(r),\,\Phi(r)$,
related through three independent equations, and consequently, there are two independent functions.

The exoticity of the required material is a result of the flaring-out condition, $b'(r)<b(r)/r$, which taking into account equations (\ref{rho}) and (\ref{pr}), implies $p_r(r)+\rho(r)<0$ at or near the throat. Thus, the stress-energy tensor violates the NEC, and consequently all the classical pointwise energy conditions. In order to reduce the exoticity of the fluid it is common to require $\rho>0$, that is $b'(r)>0$. It must be noted that the geometry described by equation~(\ref{metrica}) for general $\Phi(r)$ and $b(r)$ is not supported by a fluid with isotropic pressures in general. That is, $p_t$ is generically different from $p_r$ close to the throat (they could be equal in the asymptotic region if one requires a given behaviour at infinity, or in particular cases) and is given by $p_r$, $p_r'$ and $\rho$ through equation~(\ref{conservacion}). In fact, the first studies of phantom wormholes \cite{Sushkov:2005kj, Lobo:2005us} extended the notion of phantom energy to inhomogeneous spherically symmetric spacetimes and considered that the pressure related to the energy density through the equation of state parameter $w$ in cosmological scenarios must be the radial pressure in these geometries; the transverse pressure can then be calculated by means of the Einstein equations. That is enough to get the inequality $p_r(r)+\rho(r)<0$ on the radial direction, which is responsible for the violation of the NEC, at or near the throat.

%%%%%%%%%%%%%%%%%%%%%%%%%%%%%%%%%%%%%%%%%%%%%%%%%%%%%%%%%%%%%%%%%%%%%%%%%%%%%%%%%%%%%%%%%
%%%%%%%%%%%%%%%%%%%%%%%%%%%%%%%%%%%%%%%%%%%%%%%%%%%%%%%%%%%%%%%%%%%%%%%%%%%%%%%%%%%%%%%%%

\subsection{Asymptotic behaviour and thin shells}\label{thinshell}

In order to study the characteristics of the wormhole, it is of particular interest to
consider asymptotically flat solutions.
This asymptotic behaviour is only possible if $\Phi(r)$ and $b(r)/r$ tend 
to zero when $r\rightarrow\infty$ (it must be noted that a constant limit for $\Phi(r)$ would
also allow us to obtain asymptotically flat solutions under time reparametrization). 
One can also consider a flat space in the asymptotic limit by 
taking into account
thin shells, that is, one can cut the original spacetime at a given hypersurface and paste it together
with an exterior spacetime leading to a boundary between the two regions with a
surface stress-energy tensor that will not vanish in general \cite{Lobo:2004rp,Lobo:2005zu}.
As is well known, the stress-energy tensor can
vanish in a region given by $r>a$, if the interior part of the geometry given by metric~(\ref{metrica}) is matched
to this vacuum spacetime at $r=a>r_0$. This procedure must be done imposing the 
Israel junction conditions \cite{c1,c2,c3,c4,c5,c6} at the boundary surface $r=a$,
which in general would lead to a junction surface containing stresses, that is, both geometries are matched
together at a thin shell. It must be noted that we are constructing
a regular geometry with an interior wormhole spacetime,
$r_0\leq r\leq a$, and a Schwarzschild 
exterior\footnote{Thus, we are pasting the region inside a given hypersurface of one spacetime with the region outside a particular  hypersurface of the other spacetime. 
The situation is different when one considers two exterior geometries to be matched to construct a thin-shell wormhole as summarized in~\cite{Nadiezhda}. In this case there 
is a relative sign flip in equations (\ref{sigma}) and (\ref{p}) between the first and second addends, as has been pointed out in~\cite{gravastar} for the gravastar case.}, 
$r_{H}<a\leq r<\infty$, where $r_H$ is the event horizon (which does not exist in the resulting geometry).

Considering that the surface stress-energy tensor may be written in terms of the surface energy density,
$\sigma$, and the surface pressure, $\cal{P}$, as $S^i_j= {\rm diag}\left(-\sigma,\,\cal{P},\,\cal{P}\right)$
and applying the Lanczos equations, the surface stresses of the thin shell, which is around the exterior of
the throat, are given by \cite{oai:arXiv.org:gr-qc/0506001}
\begin{equation}\label{sigma}
 \sigma=-\frac{1}{4\pi a}\left(\sqrt{1-\frac{2M}{a}+\dot{a}^2}-\sqrt{1-\frac{b(a)}{a}+\dot{a}^2}\right)\,,
\end{equation}
\begin{equation}\label{p}
 {\cal P}=\frac{1}{8\pi a}\left[\frac{1-\frac{M}{a}+\dot{a}^2+a\ddot{a}}{\sqrt{1-\frac{2M}{a}+\dot{a}^2}}
-(1+a\,\Phi')\sqrt{1-\frac{b(a)}{a}+\dot{a}^2}+
\frac{a\ddot{a}-\frac{\dot{a}^2[b(a)-a\,b'(a)]}{2[a-b(a)]}}{\sqrt{1-\frac{b(a)}{a}+\dot{a}^2}}\right] \,,
\end{equation}
respectively, with $\dot{a}\equiv da/d\tau$, where $\tau$ is the proper time of the shell; thus $\tau\neq t$ 
\cite{Lobo:2005zu,Nadiezhda}.
There is also a flux term corresponding to the net discontinuity in the momentum flux, which leads to
a work term $\Xi$ in the conservation equation on the shell 
\begin{equation}\label{conseveq}
 \sigma'=-\frac{2}{a}\left(\sigma+{\cal P}\right)+\Xi,
\end{equation}
with
\begin{equation}\label{Xi}
 \Xi=-\frac{1}{4\pi a^2}\left[\frac{a\,b'(a)-b(a)}{2[a-b(a)]}+a\,\Phi'\right]\sqrt{1-\frac{b(a)}{a}+\dot{a}^2} \,.
\end{equation}

As we will show below, it is useful to retain the terms $\dot{a}$ and $\ddot{a}$ in equations (\ref{sigma}), (\ref{p}) 
and (\ref{Xi})  
to study the stability regions under radial perturbations. Nevertheless, we are interested in static
solutions. Substituting $\dot{a}=0$ in equations (\ref{sigma}), (\ref{p}) and (\ref{Xi}), it is immediate to see
that a static solution at $a=a_0$ leads to a thin shell characterized by:
\begin{equation}\label{sigma0}
 \sigma(a_0)=-\frac{1}{4\pi a}\left(\sqrt{1-\frac{2M}{a}}-\sqrt{1-\frac{b(a)}{a}}\right),
\end{equation}
\begin{equation}\label{p0}
 {\cal P}(a_0)=\frac{1}{8\pi a}\left[\frac{1-\frac{M}{a}}{\sqrt{1-\frac{2M}{a}}}
-(1+a\,\Phi')\sqrt{1-\frac{b(a)}{a}}\right],
\end{equation}
and
\begin{equation}\label{Xi0}
 \Xi(a_0)=-\frac{1}{4\pi a^2}\left[\frac{a\,b'(a)-b(a)}{2[a-b(a)]}+a\,\Phi'\right]\sqrt{1-\frac{b(a)}{a}}\,.
\end{equation}
A necessary requirement for the absence of horizons is $a>2\,M$. In addition, one could require:
(i) $\sigma\geq0$ on the shell, then $2\,M\geq b(a)$; and (ii) a non-negative energy density in
the interior space, then $b'(a)\geq0$, which leads to a non-decreasing $b(r)$
and $b(a)\geq b(r_0)=r_0$. Therefore, in this case $r_0\leq b(a)\leq 2\,M<a$.

In order to study the stability around a static solution, one can obtain $\dot{a}^2$ from equation (\ref{sigma}),
and re-write the resulting expression 
as \cite{Visser:2003ge,Lobo:2005zu,Nadiezhda}:
\begin{equation}\label{motion}
 \frac{1}{2}\dot{a}^2+V(a)=0,
\end{equation}
with
\begin{equation}
 V(a)=\frac{1}{2}\left\{1-\frac{\widebar{b}(a)}{a}-\left[\frac{m_s(a)}{2\,a}\right]^2-\left[\frac{\Delta(a)}{m_s(a)}\right]^2\right\},
\end{equation}
where
\begin{equation}\label{ms}
 \widebar{b}(a)=\frac{2\,M+b(a)}{2}, \qquad \Delta(a)=\frac{2\,M-b(a)}{2}, \qquad m_s(a)=4\pi\sigma(a) a^2 \,,
\end{equation}
has been defined for notational simplicity.

Linearizing the potential around a static solution at $a_0$, that is $V(a_0)=0$ and $V'(a_0)=0$, one has
\begin{equation}\label{linearized}
 V(a)=\frac{V''(a_0)}{2}(a-a_0)^2+{\cal O}\left[(a-a_0)^3\right] \,.
\end{equation}
A static solution is stable if $V''(a_0)>0$. Now, one can easily study the stability
to linearized spherically symmetric perturbations around static solutions in terms of the physical quantities
by considering the effect of this condition in the surface stress-energy tensor. As we have three functions,
given by (\ref{sigma}), (\ref{p}) and (\ref{Xi}), which are related through equation (\ref{conseveq}),
it is enough to consider the behaviour of two of these quantities under perturbations.
Thus, substituting (\ref{motion}) and (\ref{linearized}) in equation (\ref{sigma}), considering the
definition of $m_s$ in equation (\ref{ms}), deriving twice with respect to $a$, and then evaluating at $a=a_0$, one
can see that $V''(a_0)>0$ if 
\begin{equation}\label{master1}
 m_s''(a_0)\geq \frac{1}{4\,a_0^3}\left\{\frac{\left[b(a_0)-a_0b'(a_0)\right]^2}{\left[1-b(a_0)/a_0\right]^{3/2}} 
-\frac{4\,M^2}{\left(1-2\,M/a_0\right)^{3/2}}\right\}+\frac{1}{2}\frac{b''(a_0)}{\sqrt{1-b(a_0)/a_0}} \,,
\end{equation}
for $\sigma(a_0)>0$, and the inequality sign would be reversed for configurations with $\sigma(a_0)<0$.
This is the first master equation. 

On the other hand, considering now the function $4\pi\,a\,\Xi$ 
and following a similar procedure, one obtains the second master equation. This is\footnote{Equation (\ref{master2})
differs from the second master equation presented in reference \cite{gravastar}, since we are expressing
the temporal component of metric (\ref{metrica}) in a different form.}:
\begin{eqnarray}\label{master2}
 \left[4\pi\,a\Xi\right]''&\leq&\frac{3[u'(a_0)]^3}{8[u(a_0)]^{5/2}}-\frac{3\,u'(a_0)u''(a_0)}{4[u(a_0)]^{3/2}}+
\frac{u'''(a_0)}{2[u(a_0)]^{1/2}}+\frac{\Phi'(a_0)}{2}\left[\frac{[u'(a_0)]^{2}}{2[u(a_0)]^{3/2}}-\frac{u''(a_0)}{[u(a_0)]^{1/2}}\right]
\nonumber\\
&&-\Phi''(a_0)\frac{u'(a_0)}{[u(a_0)]^{1/2}}-\Phi'''(a_0)[u(a_0)]^{1/2},
\end{eqnarray}
for
\begin{equation}\label{cond2}
 \frac{u'(a_0)}{2\,u(a_0)}-\Phi'(a_0)>0,
\end{equation}
where we have defined
\begin{equation}
 u(a)=1-\frac{b(a)}{a}.
\end{equation}
It must be noted that the sign will be reversed in equation (\ref{cond2}) otherwise. We have chosen to express
the second master equation with the inequality as expressed in (\ref{master2}), that is, when condition 
(\ref{cond2}) is satisfied, because the latter condition reduces to the flaring-out condition for
$\Phi'(a_0)=0$. As we will show, one can conclude in which case the configuration is more stable studying
the parameters which increases the stability region bounded by the surfaces given by (\ref{master1})
and (\ref{master2}).

%%%%%%%%%%%%%%%%%%%%%%%%%%%%%%%%%%%%%%%%%%%%%%%%%%%%%%%%%%%%%%%%%%%%%%%%%%%%%%%%%%%%%%%%%
%%%%%%%%%%%%%%%%%%%%%%%%%%%%%%%%%%%%%%%%%%%%%%%%%%%%%%%%%%%%%%%%%%%%%%%%%%%%%%%%%%%%%%%%%

\section{Constant minimal violations of the NEC}\label{constantes}

\subsection{Equation of state and general considerations}

Let us now consider our particular model to restrict the form of the redshift and shape functions. 
The equation of state minimizing the violation of the NEC that we will consider is inspired by the equation of state presented in reference~\cite{Mariam}, where the cosmological implications of a
small constant deviation from a cosmological constant were investigated; that is, a cosmological model with $\rho(t)+p(t)=-A$, where
the parameter $A$ must be small compared with the parameters of the model \cite{Mariam}.  One can assume that in less symmetric situations such equation of state would generically take the form
\begin{equation}\label{generic}
 \rho\left(x^\mu\right)+p\left(x^\mu\right)=-A.
\end{equation}
As is well-known, this fluid description can also be interpreted as following from a field model. In particular, one can consider the equivalence of a barotropic perfect fluid with a scalar field
model with an action which is a function of the field $\phi$ and its kinetic term $X=-1/2\,g^{\mu\nu}\partial_\mu\phi\,\partial_\nu\phi$ (k-essence field models)
\cite{ArmendarizPicon:1999rj,Arroja:2010wy}. That is, for a field action of the form
\begin{equation}
 S_\phi=\int {\rm d}^4 x \sqrt{-g}\,K(X,\,\phi),
\end{equation}
assuming that $\partial_\mu\phi$ is a time-like vector, the field stress energy tensor takes the form of a perfect fluid with  \cite{ArmendarizPicon:1999rj,Arroja:2010wy}
\begin{equation}\label{tranfor}
 \rho=2X\,K_{,X}-K,\qquad p=K, \qquad u_\mu=\frac{\partial_\mu\phi}{\sqrt{2X}}.
\end{equation}
Therefore, if $K_{,X}$ is not constant, compatibility between equations (\ref{generic}) and (\ref{tranfor}) fixes the particular field model to be described by
\begin{equation}
 K(X,\,\phi)=-\frac{A}{2}{\rm ln}\left(X\right)-V(\phi),
\end{equation}
where the potential appears as an integration constant. Choosing a particular function $V(\phi)$ implies fixing the form the functions $\rho(x^\mu)$ and $p(x^\mu)$ satisfying
equation (\ref{generic}). We prefer, however, these functions to be fixed by considering particular models of interest, therefore, we leave $V(\phi)$ arbitrary for the moment.
On the other hand, a constant $K_{,X}$, which corresponds to a 
canonical kinetic term for the scalar field, would instead lead to a phantom model with a constant kinetic term.

In our case we consider an anisotropic generalization of equation (\ref{generic}) suitable for a wormhole geometry, as it is usually done in the literature \cite{Sushkov:2005kj,Lobo:2005us}. 
It must be pointed out that such an anisotropic generalization would need to go beyond the k-essence field theoretical model \cite{Korolev:2014hwa}.
This is
\begin{equation}\label{estado}
 \rho(r)+p_r(r)=-\frac{A}{8\pi},
\end{equation}
where the $8\pi$ has been introduced for convenience and the functions only depend on the radial coordinate due to spherical symmetry. 
In order for the NEC to be violated only in a minimal way, we have to demand $A$ to be small compared with the parameters of the model.
For simplicity and by dimensional analysis we consider
\begin{equation}\label{pequeno}
 A<1/r_0^2,
\end{equation}
although it should be pointed out that for values of $A$ small enough other more restrictive bounds could also be satisfied. 

We note that inequality (\ref{pequeno}) can be obtained by considering a barotropic equation of state given by $p_r=\omega \rho$, with $\omega={\rm const}<-1$ \cite{Sushkov:2005kj}. For this case, the radial pressure at the throat is given by $p_r|_{r_0}=-1/(8\pi r_0^2)$, so that the energy density at $r_0$ is provided by $\rho |_{r_0}=-1/(8\pi \omega r_0^2)<1/(8\pi r_0^2)$. Now, taking into account $p_r=\omega \rho$ and using equation $(\ref{estado})$, we find that the energy density is given by $\rho=-A/[8\pi (1+\omega)]$. Using the latter expression and taking into account $\rho |_{r_0}=-1/(8\pi \omega r_0^2)$, we find $A=|1+\omega |/(|\omega | r_0^2)$, from which inequality (\ref{pequeno}) automatically follows.

Considering the barotropic equation of state considered above, i.e., $p_r=\omega \rho$, for the specific case of $\omega={\rm const}<-1$, 
the energy density $\rho=-A/[8\pi (1+\omega)]$ is constant throughout the spacetime. We emphasize that the constant energy density was analysed in \cite{Sushkov:2005kj}, 
but for different redshift functions than the ones we consider in this work. Nevertheless, it must be pointed out that in this paper we are not restricting our attention to the case of a constant equation
of state parameter $w$.
Furthermore, we also extend the analysis to the dynamical stability of the thin shell separating the interior and exterior regions.

Taking into account the equation of state (\ref{estado}) and the Einstein equations (\ref{rho}) and 
(\ref{pr}), we obtain a relation between the redshift and shape function, given by
\begin{equation}\label{master}
 \left(1-\frac{b}{r}\right)\Phi'=\frac{r}{2}\left(-A+\frac{b-b'r}{r^3}\right).
\end{equation}
Evaluating this expression at $r_0$ one gets
\begin{equation}
 A=\frac{1-b'(r_0)}{r_0^2}.
\end{equation}
Thus, as it could be expected from the general analysis, $A>0$ to satisfy the flaring-out condition
($b'(r_0)<1$).
Moreover, a minimal condition for satisfying the requirement of minimal violations, expressed
in~(\ref{pequeno}), is $b'(r_0)>0$; this condition implies $\rho(r_0)>0$ through equation (\ref{rho}), 
which is a desirable property. Thus, once one fixes a particular shape function, the redshift function
can be obtained by integrating equation (\ref{master}). On the other hand, defining $u(r)=1-b(r)/r$ one can 
re-write equation (\ref{master}) as
\begin{equation}\label{masterb}
 u'(r)-2\,\Phi'(r)\,u(r)-A\,r=0,
\end{equation}
which can be easily solved and leads to
\begin{equation}\label{bconstant}
b(r)=r\left[1-A\,e^{2\Phi(r)} \int_{r_0}^r dr'\, r' e^{-2\Phi(r')}\right],
\end{equation}
where the integration constant was fixed taking into account the condition $b(r_0)=r_0$.

On the other hand, it can be noted that it is not possible to have asymptotically flat solutions if the
whole space is described by metric (\ref{metrica}).
This is because $\rho+p_r$ is equal to a 
constant value even in the asymptotic limit, whereas one would need to have $\rho+p_r\rightarrow0$
when $r\rightarrow\infty$ to have $\Phi(r)\rightarrow0$ and $b(r)/r\rightarrow 0$, that is,
an asymptotically flat geometry (or asymptotically de Sitter or
anti de Sitter). 
Thus, in order to construct asymptotically flat geometries, it is necessary for the equation of state, given by equation (\ref{estado}), that the interior wormhole is surrounded by a thin shell.
In the following we will consider some particular wormhole solutions with the phantom fluid described 
by the equation of state (\ref{estado}).

%%%%%%%%%%%%%%%%%%%%%%%%%%%%%%%%%%%%%%%%%%%%%%%%%%%%%%%%%%%%%%%%%%%%%%%%%%%%%%%%%%%%%%%%%
%%%%%%%%%%%%%%%%%%%%%%%%%%%%%%%%%%%%%%%%%%%%%%%%%%%%%%%%%%%%%%%%%%%%%%%%%%%%%%%%%%%%%%%%%

\subsection{Constant redshift function}\label{constantphi}

Let us consider $\Phi=\Phi_0$, then the two exponentials of equation (\ref{bconstant}) can be simplified, and  one gets
\begin{equation}
 b(r)=-\frac{A}{2}r^3+\left(\frac{A\,r_0^2}{2}+1\right) r,
\end{equation}
where the integration constant has been fixed noting that $b(r_0)=r_0$, leading to a result independent of $\Phi_0$. The line element (\ref{metrica}) can then be written as
\begin{equation}\label{metrichom}
 ds^2=-e^{2\Phi_0}dt^2+\frac{2}{A}\frac{dr^2}{r^2-r_0^2}+r^2 d\Omega_{(2)}^2,
\end{equation}
where $d\Omega_{(2)}^2=(d\theta^2+ \sin^2d\varphi^2)$.

\begin{figure}[h]
\centering
\includegraphics{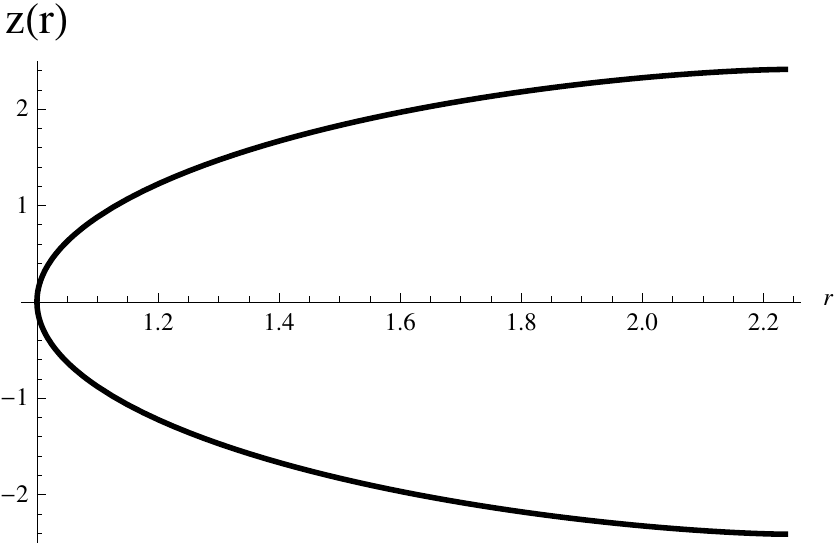}
\caption{Embedded surface for $r_0=1$ and $A=1/2$. See the text for more details.}
\label{fig1}
\end{figure}

Moreover, one can calculate the proper distance (\ref{lr}), given by
\begin{equation}
 l(r)=\pm \sqrt{\frac{2}{A}}{\rm ln}\left(r/r_0+\sqrt{r^2/r_0^2-1}\right),
\end{equation}
which is well-defined in the whole space. One can embed the wormhole in one extra dimension close to
the throat. The embedded surface $z(r)$ can be calculated integrating equation (\ref{zr}) and yields
\begin{eqnarray}
 z(r)=\pm\int^r_{r_0}dr\left(\frac{r_0^2+\frac{2}{A}-r'^2}{r'^2-r_0^2}\right)^{1/2},
\end{eqnarray}
which is well-defined only close to the throat, that is, for $r_0<r<\sqrt{(2+A\,r_0^2)/A}$. Integrating this expression we get \cite{Gradshteyn}
\begin{eqnarray}
 z(r)&=&
\pm \sqrt{r_0^2+\frac{2}{A}}\left[F\left({\rm arcsin}\sqrt{\left(1+\frac{A}{2}r_0^2\right)\left(1-\frac{r_0^2}{r^2}\right)}
,\,\left(1+\frac{A}{2}r_0^2\right)\right)\right.\nonumber\\
&&-\left.E\left({\rm arcsin}\sqrt{\left(1+\frac{A}{2}r_0^2\right)\left(1-\frac{r_0^2}{r^2}\right)}
,\,\left(1+\frac{A}{2}r_0^2\right)\right)\right]
\nonumber\\  &&
+\frac{1}{r}\sqrt{\left(r_0^2+\frac{2}{A}-r^2\right)\left(r^2-r_0^2\right)},
\end{eqnarray}
where $E(\phi,\,k)$ and $F(\phi,\,k)$  are the elliptic integral of the second kind and the elliptic integral of the first kind, respectively.
The embedded function, is depicted in figure \ref{fig1}, for a particular wormhole solution, choosing the values for $r_0=1$ and $A=1/2$ for the parameters. Figure \ref{fig1} illustrates how this surface flares out in the additional auxiliary dimension in which $z$ is defined.

Regarding the exotic fluid some comments are in order. From the equations (\ref{rho})--(\ref{pt}), 
one obtains
\begin{equation}\label{rho1}
 \rho=\frac{1}{8\pi}\left(-\frac{3\,A}{2}+\frac{1+\frac{A}{2} r_0^2}{r^2}\right),
\end{equation}
\begin{equation}\label{pr1}
 p_r=\frac{1}{8\pi}\left(\frac{A}{2}-\frac{1+\frac{A}{2}r_0^2}{r^2}\right),
\end{equation}
\begin{equation}\label{pt1}
 p_t=\frac{A}{16\pi},
\end{equation}
respectively, and of course, equation (\ref{estado}) is satisfied.
Thus, $p_t$ is equal to a positive constant throughout the whole space. 
Close to the wormhole throat one has 
$\rho>0$ and $p_r<0$. Then, the energy density vanishes at $r_*=\sqrt{(2+A\,r_0^2)/(3\,A)}$, and becomes negative for $r>r_*$. The radial pressure becomes zero at $r_{**}=\sqrt{(2+A\,r_0^2)/A}$, and negative for $r>r_{**}$. 
As the metric (\ref{metrichom}) is not well-defined in the asymptotic limit, one matches this interior wormhole solution to an exterior Schwarzschild spacetime. It must be noted that it is of specific interest to consider a shell 
with $a_0< r_*$ to avoid negative energy density (\ref{rho1}). 
Thus, in this case, the QNEC will be satisfied since one would have $\rho(r)+p_t(r)>0$ and 
$\rho(r)+p_r(r)=-A/(8\pi)$ inside the shell, and these quantities vanishing outside.
From equations (\ref{sigma0})--(\ref{Xi0}), such a static shell is characterized by the following surface stresses
\begin{equation}\label{sigma11}
 \sigma (a_0)=\frac{1}{4\pi a_0}\left(\sqrt{\frac{A}{2}}\sqrt{a_0^2-r_0^2}-\sqrt{1-\frac{2\,M}{a_0}}\right),
\end{equation}
\begin{equation}\label{P11}
 {\cal P}(a_0)=\frac{1}{8\pi a_0}\left(\frac{1-\frac{M}{a_0}}{\sqrt{1-\frac{2\,M}{a_0}}}-
\sqrt{\frac{A}{2}}\sqrt{a_0^2-r_0^2}\right),
\end{equation}
and
\begin{equation}\label{Xi11}
 \Xi(a_0)=\frac{1}{4\pi}\sqrt{\frac{A}{2}}\frac{1}{\sqrt{a_0^2-r_0^2}},
\end{equation}
respectively, where $r_0\leq 2M<a_0<r_*$. It must be noted that if one considers the shell placed at $a_0=r_*$  to avoid discontinuities in the function $\rho$ through the bulk space (since $\rho$ vanishes there) one would still have a non-vanishing $\sigma$ on the shell because the bulk pressure is not vanishing at this radial coordinate. In fact, it can be seen that $\sigma>0$ for $0<r_0<2\,M<a_0$. 

\begin{figure}[h]
\centering
\includegraphics{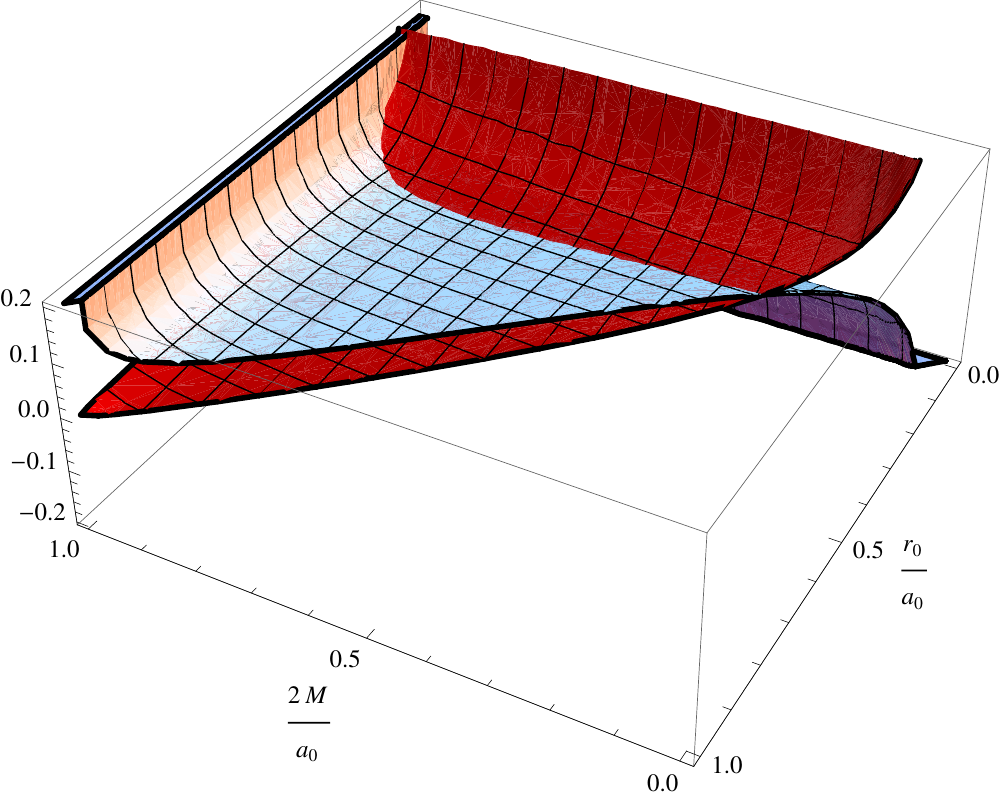}
\caption{$a_0\,\sigma$ (lower surface for large values of $2M/a_0$ and of $r_0/a_0$) and $a_0\,{\cal P}$ (upper surface for large values of $2M/a_0$ and of $r_0/a_0$) are depicted for  $A\,r_0^2=1/10$, in the range $0<2M/a_0<1$ and $0<r_0/a_0<2M/a_0$. The function $a_0\,\sigma$ is positive in the whole interval, and the  function $a_0\,{\cal P}$ is negative for small values of $r_0/a_0$ and positive for large values of this quantity.}
\label{fig2}
\end{figure}

Defining $x= 2M/a_0$ and $y=r_0/a_0$, then $0<x<1$, $0<y<x$, and indeed $\sqrt{\frac{3Ar_0^2}{2+Ar_0^2}}<y<x$ to avoid negative values for the energy density in the interior region (\ref{rho1}), one can rewrite equations (\ref{sigma11})--(\ref{Xi11}), as
\begin{equation}\label{sigmahom}
 \sigma=\frac{1}{4\pi a_0}\left(\sqrt{\frac{A}{2}}\,r_0\sqrt{y^{-2}-1}-\sqrt{1-x}\right),
\end{equation}
\begin{equation}
 {\cal P}=\frac{1}{8\pi a_0}\left(\frac{1-\frac{x}{2}}{\sqrt{1-x}}-
\sqrt{\frac{A}{2}}\,r_0\sqrt{y^{-2}-1}\right),
\end{equation}
and
\begin{equation}
 \Xi(a_0)=\frac{1}{4\pi r_0}\sqrt{\frac{A}{2}}\frac{y}{\sqrt{1-y^2}},
\end{equation}
respectively, where $r_0\sqrt{A/2}<\sqrt{1/2}$. Thus, one can plot $a_0\,\sigma$ and $a_0\,{\cal P}$ for a fixed value of $r_0\sqrt{A/2}$ to understand the behaviour of the surface stress-energy tensor, which are depicted in figure \ref{fig2}. It can be seen that whereas $\sigma$ is larger than zero, ${\cal P}$ could be less than zero for small values of $y$.

\begin{figure}[h]
\centering
\includegraphics{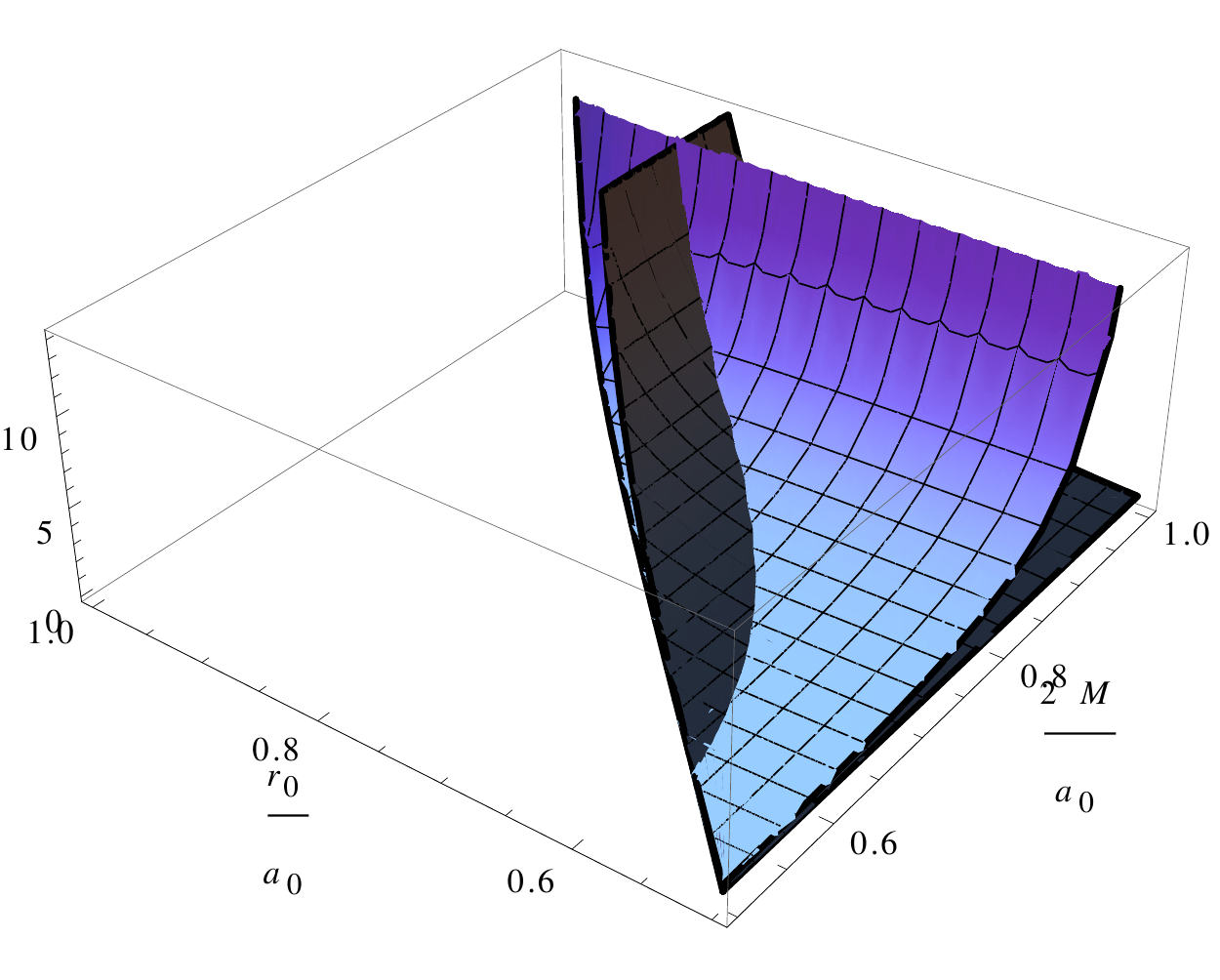}
\caption{
Case $Ar_0^2=1/10$. The surface corresponding to the first master equation (\ref{m11}) and
the surface associated to the second master equation (darker colour) (\ref{mas12})
are depicted, where we show only the
region corresponding to positive energy density in the interior region, given by equation (\ref{rho1}). 
A stable configuration should have $ a_0 \,m''_s(a_0)$ above the clearer surface and $ a_0^3\,(4\pi a \Xi)''$ below the darker surface. Thus, the final stability region is the region bounded by both surfaces. }
\label{fig3}
\end{figure}

In order to study the stability of the shell, we consider, in first place, the first master equation.
The inequality (\ref{master1}) can be expressed for this wormhole solution as
\begin{equation}\label{m11}
 a_0 \,m''_s(a_0)\geq \left[\frac{x^2}{4(1-x)^{3/2}}+\sqrt{\frac{A}{2}}\,r_0\frac{2-3y^2}{y(1-y^2)^{3/2}}\right]
\end{equation}
Thus, it can be easily noted that for smaller values of $r_0\sqrt{A}$ the r.h.s. of equation (\ref{m11}) is larger
for $y<\sqrt{2/3}$,
thus the stability region would be smaller; whereas the stability region would be larger for 
smaller values of $r_0\sqrt{A}$ if $y>\sqrt{2/3}$. In figure \ref{fig3} one can see that 
the stability region is larger for smaller values of $r_0/a_0$ and values $2\,M/a_0$ not too close
to one (it can be noted that for different values of $r_0\sqrt{A}$ the variation of $2\,M/a_0$ 
does not lead to a large variation on the surface position, until this quantity is close to one).

Next, we consider the second master equation (\ref{master2}), which reduces to
\begin{equation}\label{mas12}
 a_0^3\,(4\pi a \Xi)''\leq 3\sqrt{\frac{A}{2}}r_0\frac{y}{(1-y^2)^{5/2}}.
\end{equation}
The stability region associated to this master equation is larger for larger values of $A$ and
$r_0/a_0$ (although it would be smaller for larger values of $A$ and smaller values of $r_0/a_0$), 
see figure \ref{fig3}. 
Therefore, taking into account both master equations, the values of $2\,M/a_0$ (which is not appearing in equation
(\ref{mas12}))
not too close to $1$ would be preferred,
thus one should avoid $2\,M\sim a_0$. 
Both surfaces diverges for $r_0/a_0\rightarrow1$, but the second surface diverges faster at this limit. Both stability regions have the same behaviour under variations of 
$A\,r_0$ if $y>\sqrt{2/3}$. Thus, one can consider $\sqrt{2/3}\,a_0\lesssim r_0<2\,M<a_0$
to maximize the stability, taking large departures with respect a cosmological constant,
$A\,r_0\lesssim1$,
as it could be expected.
As stable configurations should have $ a_0 \,m''_s(a_0)$ above the first surface and $ a_0^3\,(4\pi a \Xi)''$ bellow the second surface, then the stable configuration would be larger in regions of the parameters $x=2\,M/a_0$ and $y=r_0/a_0$ for which the first surface reaches larger values and the second surface (the darker surface
in figure~\ref{fig3}) takes small values. Both master equations are expressed using dimensionless quantities, and the final stability regions are bounded by both surfaces represented by the master equations. 
More specifically, configurations with $x,\,y$ in the region where the first surface is below the second one depict the final stability region \cite{Nadiezhda}.

\section{Inhomogeneous minimal violations of the NEC}\label{inhomogeneous}

The equation of state (\ref{estado}), considered in the previous section, is not compatible
with an asymptotically flat limit due to the strong constraint of a constant value of $A$. This constraint is
not so stringent in a cosmological scenario \cite{Mariam}, where it could be thought to be even natural. 
Nevertheless, one
could consider that in order to study astrophysical objects one should not only assume a radial dependence
of the l.h.s. of equation (\ref{estado}), but one has also to take into account a dependence of the
r.h.s. on the radial coordinate. Assuming that the equation of state should also be compatible with 
an asymptotic vacuum regime,
the simplest equation that one can write is:
\begin{equation}\label{inh}
 \rho+p_r=-\frac{A}{8\pi}\left(\frac{r_0}{r}\right)^\alpha,
\end{equation}
with $\alpha>0$ a constant parameter. 
Following the spirit drawn in the introduction, we consider that the violation of the NEC has to be minimal
also in this case. Thus, we again impose 
\begin{equation}\label{ineq2}
 A<1/r_0^2.
\end{equation}

As emphasized in the previous section, the inequality (\ref{ineq2}) may be deduced by considering a barotropic equation of state $p_r=\omega \rho$, with $\omega={\rm const}<-1$. 
For this case confronting the latter with equation (\ref{inh}), we deduce $\rho(r)=-\frac{A(r)}{8\pi (1+\omega)}$, where $A(r)=A_0(r_0/r)^{\alpha}$. 
It is interesting to note that in reference~\cite{Sushkov:2005kj} the specific case of an energy density following a normal Gaussian distribution law, given by $\rho(r)=\rho_0 e^{\bar{\alpha} (r/r_0-1)^2}$ 
where $\bar{\alpha}>0$ is a model parameter and $\rho_0=-(8\pi \omega r_0^2)^{-1}$ is the value of the energy density at the throat, was considered. 
For this specific case, we note that $A(r)$ is given by $A(r)=\frac{1+\omega}{\omega}\,e^{\bar{\alpha} (r/r_0-1)^2}$, which is different that the solutions that we consider in this section.

It can be noted that the equation of state (\ref{inh}) is in more agreement with our spirit of minimal
violations of the NEC than equation (\ref{estado}), since
\begin{eqnarray}
 \rho+p_r\rightarrow -A,\qquad {\rm when}\qquad r\rightarrow r_0,
\nonumber\\
 \rho+p_r\rightarrow 0,\qquad  {\rm when}\qquad r\rightarrow\infty.
\end{eqnarray}
Therefore, the NEC would be marginally satisfied in the asymptotic limit (if we also have $p_t\rightarrow0$).

Considering the equation of state (\ref{inh}), the Einstein equations (\ref{rho}) and (\ref{pr}) yield
\begin{equation}\label{masterinh}
 \Phi'=\frac{r}{2\left(1-b/r\right)}\left[-A\left(\frac{r_0}{r}\right)^\alpha+\frac{b-b'r}{r^3}\right].
\end{equation}
Thus, if we consider a particular shape function, this equation can be integrated to obtain $\Phi(r)$.
On the other hand, as in the previous case we can write equation (\ref{masterinh}) in terms of $u=1-b(r)/r$ to get
the differential equation
\begin{equation}\label{masterbinh}
 u'(r)-2\,\Phi'(r)\,u(r)-A\,r_0^\alpha\, r^{1-\alpha}=0.
\end{equation}
Note that equation (\ref{masterb}) can be obtained by substituting $\alpha=0$ in equation 
(\ref{masterbinh}), as expected. The solution of equation (\ref{masterbinh}) is given by
\begin{equation}\label{binh}
b(r)=r\left[1-A\,r_0^\alpha \,e^{2\Phi(r)} \int_{r_0}^r dr'\, (r')^{1-\alpha} 
e^{-2\Phi(r')}]\right].
\end{equation}

%%%%%%%%%%%%%%%%%%%%%%%%%%%%%%%%%%%%%%%%%%%%%%%%%%%%%%%%%%%%%%%%%%%%%%%%%%%%%%%%%%%%%%%%%
%%%%%%%%%%%%%%%%%%%%%%%%%%%%%%%%%%%%%%%%%%%%%%%%%%%%%%%%%%%%%%%%%%%%%%%%%%%%%%%%%%%%%%%%%

\subsection{Shape function $b(r)=r_0(r/r_0)^\beta$}\label{inhshape}

Let us fix a shape function compatible with an asymptotically flat regime
to look for solutions of equation (\ref{masterinh}) compatible with 
a finite $\Phi(r)$ for any value of $r\geq r_0$. We choose the shape function introduced in
reference \cite{Lobo:2005us} for that purpose, that is
\begin{equation}\label{bin1}
 b(r)=r_0(r/r_0)^\beta ,
\end{equation}
where $0<\beta<1$, in order to have $0<b'(r_0)<1$ and $b(r)/r \rightarrow 0$ as $r \rightarrow \infty$. Substituting this function and its derivatives in equation (\ref{masterinh}), one gets
\begin{equation}
 \Phi(r)=\int_{r_0}^r\frac{1}{2\,r'\left[1-(r_0/r')^\gamma\right]}\left[\gamma (r_0/r')^\gamma-\frac{A}{(r')^2}(r_0/r')^\alpha\right],
\end{equation}
with $\gamma=1-\beta$, then $0<\gamma<1$, and $\alpha>0$.
It can be seen that the integral generically diverges at $r_0$, which implies that it is
not a wormhole solution. In fact, there is only one case in which
this integral converges at $r_0$, namely, for
\begin{equation}
 \alpha=2,\qquad {\rm and}\qquad \gamma=A\,r_0^2<1.
\end{equation}

\begin{figure}[h]
\centering
\includegraphics{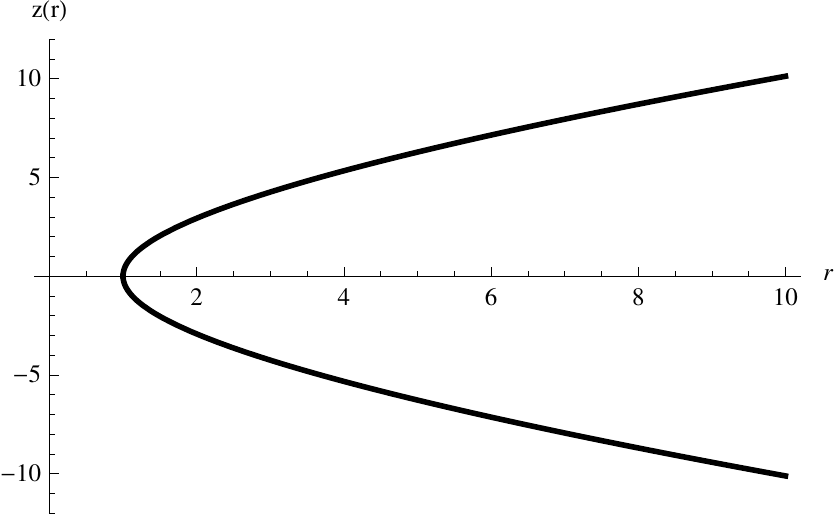}
\caption{Embedded surface for $\gamma=1/2$. See the text for details.}
\label{fig4}
\end{figure}

In this case one has
\begin{equation}\label{Phiin1}
 \Phi(r)=\frac{A}{2}r_0^2\,{\rm ln}\left(\frac{r_0}{r}\right),
\end{equation}
which diverges in the asymptotic limit, so that this solution is not asymptotic flat. The metric can be written as
\begin{equation}\label{metricain1}
 ds^2=-\left(\frac{r_0}{r}\right)^{\gamma}dt^2+\frac{dr^2}{1-(r_0/r)^\gamma}+r^2 d\Omega_{(2)}^2,
\end{equation}
with $\gamma=A\,r_0^2$.
The proper distance (\ref{lr}) is given in terms of hypergeometric functions as \cite{Gradshteyn}
\begin{eqnarray}
 l(r)&=&\pm\int_{r_0}^r\frac{dr}{\sqrt{1-(r_0/r)^\gamma}}
    \nonumber \\ 
& =& \pm \frac{2r_0}{\gamma}\sqrt{1-(r_0/r)^\gamma}\,
F\left(\frac{1}{\gamma}+1,\,\frac{1}{2};\,\frac{3}{2};\,\sqrt{1-(r_0/r)^\gamma}\right) ,
\end{eqnarray}
where $F(\alpha,\,\beta;\,\gamma;\,z)$ is the Gauss hypergeometric function,
which is well defined for\footnote{A hypergeometric
series $\textrm{F}(b,c;d;e)$, also called a hypergeometric
function, converges at any value $e$ such that $|e|\leq 1$,
whenever $b+c-d<0$. However, if  $0 \leq  b+c-d < 1$ the series
does not converge at $e=1$. In addition, if  $1 \leq b+c-d$, the
hypergeometric function blows up at $|e|=1$
\cite{Abramowitz}.} $0<\gamma<1$ and diverges for $r\rightarrow\infty$. 
The embedded surface (\ref{zr}) is given by \cite{Gradshteyn}
\begin{eqnarray}
 z(r)&=&\pm\int_{r_0}^r\frac{dr}{\sqrt{(r/r_0)^\gamma-1}}
    \nonumber \\ 
& = &
\pm \frac{2r_0}{\gamma}\sqrt{(r/r_0)^\gamma-1}\,
F\left(\frac{1}{\gamma}-1,\,\frac{1}{2};\,\frac{3}{2};\,-\sqrt{(r/r_0)^\gamma-1}\right),
\end{eqnarray}
which is again well-defined. A particular case is shown in figure \ref{fig4}, for $\gamma =1/2$.

Taking into account equations (\ref{bin1}) and (\ref{Phiin1}), with $\gamma=1-\beta$, in equations (\ref{rho})--(\ref{pt}), one gets
\begin{equation}
\rho(r)=\frac{1-\gamma}{8\pi r^2}\left(\frac{r_0}{r}\right)^\gamma ,
\end{equation}
\begin{equation}
 p_r(r)=-\frac{1}{8\pi r^2}\left[(1-\gamma)\left(\frac{r_0}{r}\right)^\gamma +\gamma\right],
\end{equation}
\begin{equation}
 p_t(r)=\frac{\gamma}{32\pi r^2}\left[2(1-\gamma)\left(\frac{r_0}{r}\right)^\gamma +\gamma\right].
\end{equation}
Therefore, as $0<\gamma<1$, the energy density is positive throughout the whole spacetime and the radial pressure is negative, and both tend to zero in the asymptotic limit by construction. 
Moreover, the transverse pressure is positive in the whole space, and it also goes to zero when 
$r\rightarrow\infty$. Therefore, the QNEC can be satisfied in the whole space for values of $A$ small enough.

\begin{figure}
\centering
\includegraphics{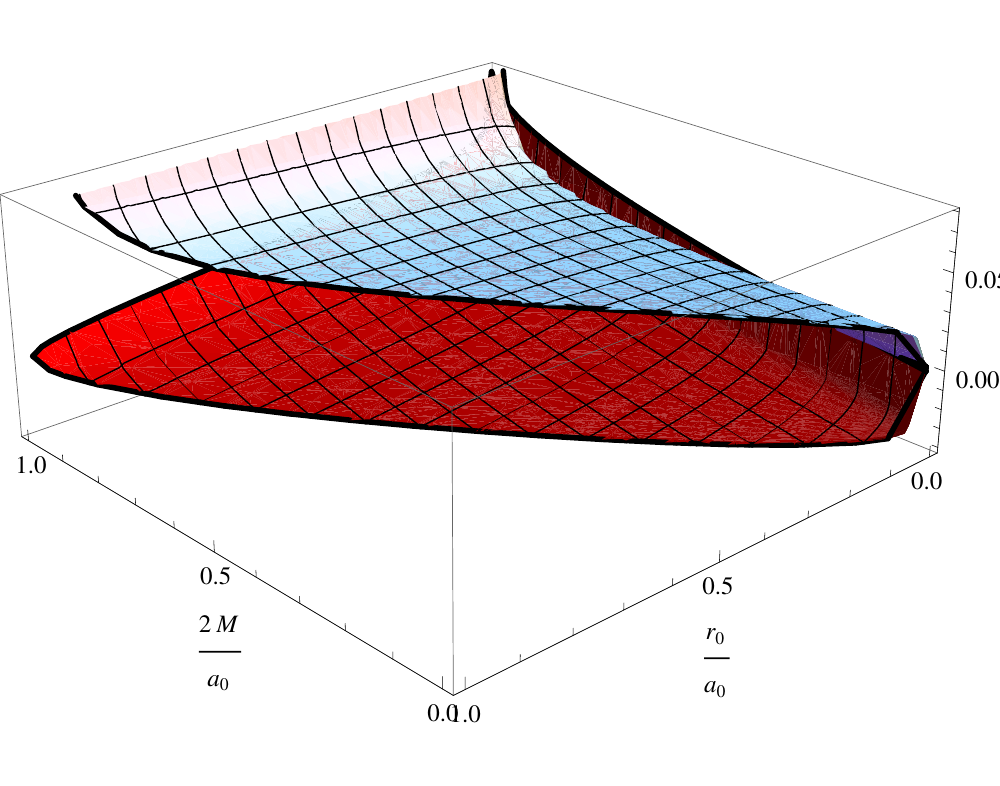}
\caption{$a_0\,\sigma$ (lower surface during most of the interval) and $a_0\,{\cal P}$
(upper surface during most of the interval) are depicted for  $\gamma=1/10$. 
The function $a_0\,\sigma$  is
positive only in a small region of the interval, whereas $a_0\,{\cal P}$ is positive in the whole region.}
\label{fig5}
\end{figure}

\begin{figure}[h]
\centering
\includegraphics{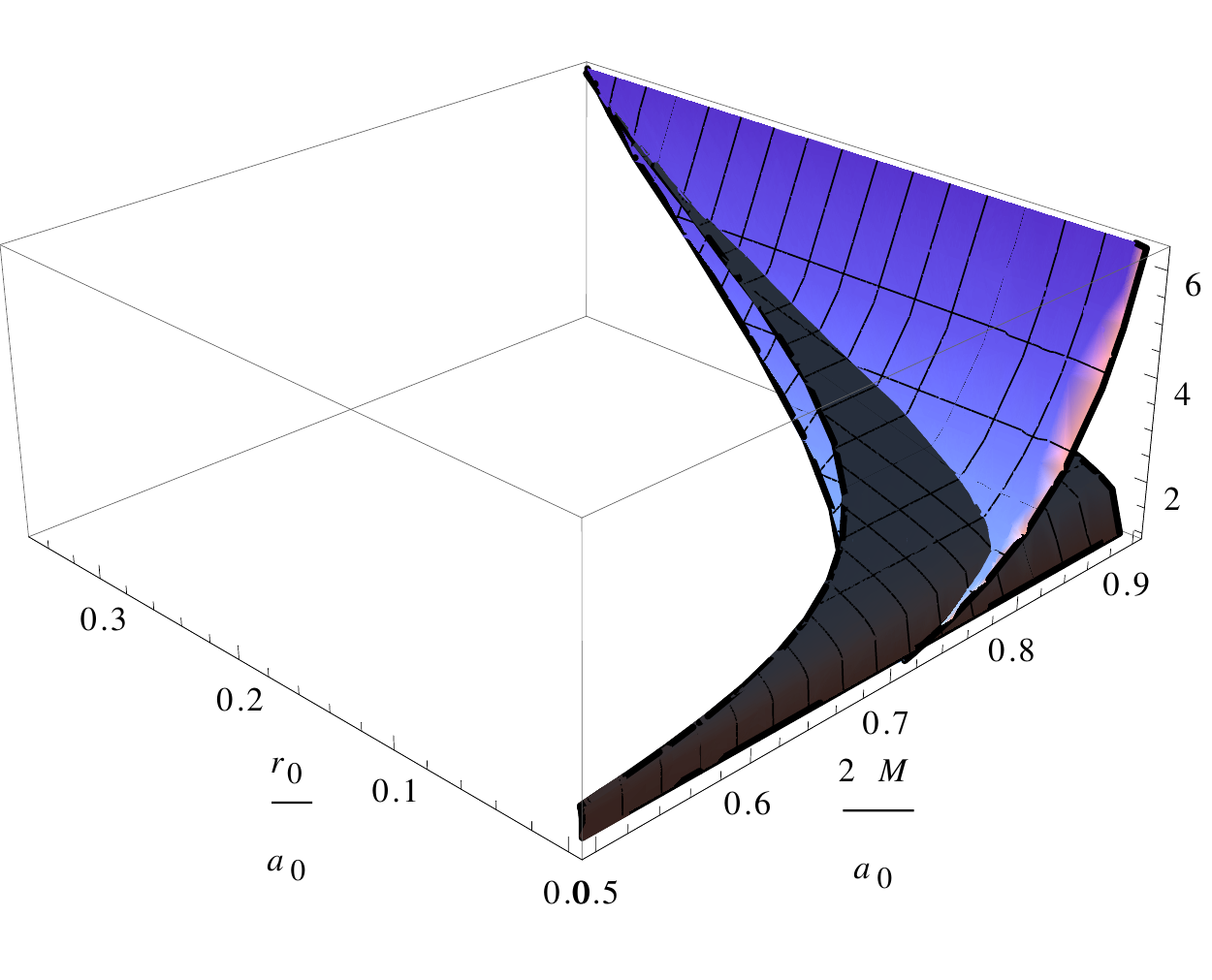}
\caption{Case $\gamma=1/10$. 
Stable configurations would have values of $a\,m''_s(a_0)$ above the clearer surface, and values of $a_0^3(4\pi a_0\Xi)''$  below the darker surface.
The stability region is shown, where 
we are considering positive values of $\sigma$. }
\label{fig6}
\end{figure}

\begin{figure}[h]
\centering
\includegraphics{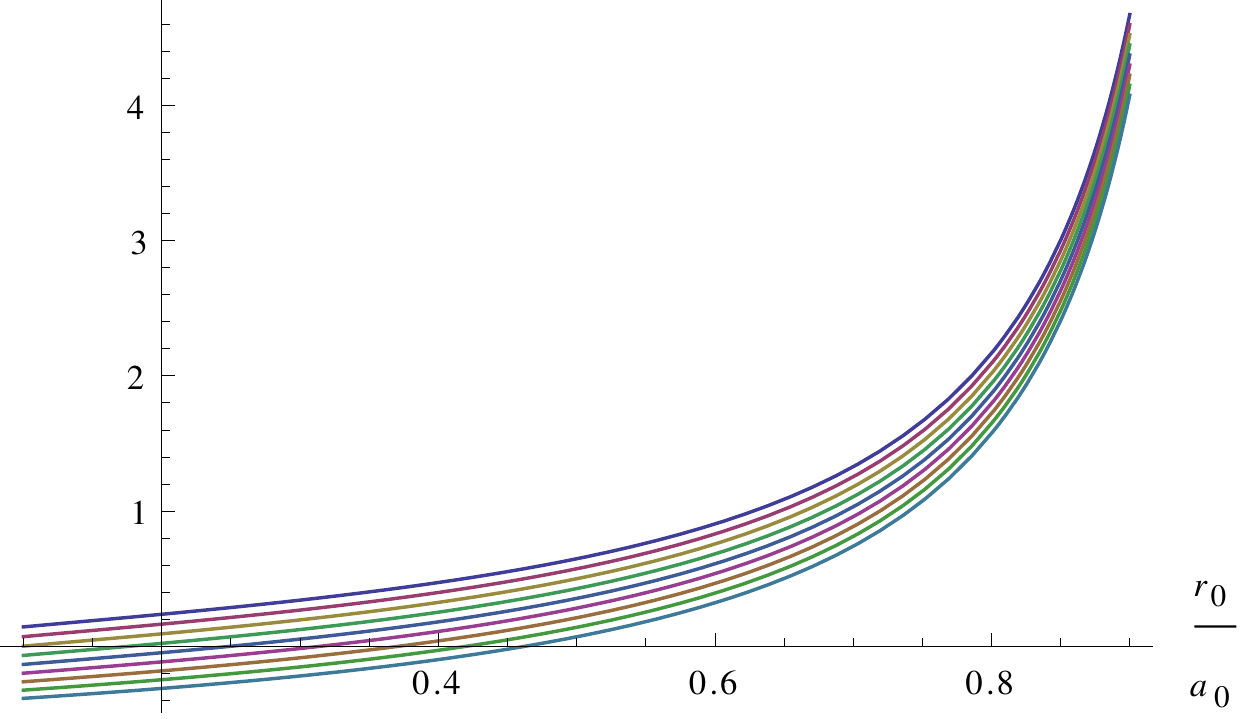} 
\caption{The condition leading to consider a particular inequality sign in the second
master equation, the inequality (\ref{minh2b}), is depicted. 
The different curves have smaller to larger values of $\gamma$ from the top to the bottom. For the value $\gamma=1/10$ shown in figure \ref{fig6}, the curve is larger
than zero for $r_0/a_0>0.001$.}
\label{fig7}
\end{figure}

As mentioned above, the metric (\ref{metricain1}) is not asymptotically flat, so let us perform the usual cut-and-paste surgery, by cutting our geometry at $r=a_0$ and pasting it to an exterior Schwarzschild geometry. 
Using
again coordinates $x=2M/a_0$ and $y=r_0/a_0$, $0<y<x$ and $0<x<1$, the physical quantities characterizing the resulting shell can be obtained using equations (\ref{sigma0})--(\ref{Xi0}), and are given by
\begin{eqnarray}\label{sigmainh}
 \sigma&=&\frac{1}{4\pi a_0}\left(\sqrt{1-y^\gamma}-\sqrt{1-x}\right), \\
 {\cal P}&=&\frac{1}{8\pi a_0}\left[\frac{1-x/2}{\sqrt{1-x}}-\left(1-\gamma/2\right)\sqrt{1-y^\gamma}\right],
\end{eqnarray}
and
\begin{equation}
 \Xi=\frac{\gamma}{8\pi a_0^2}\frac{1}{\sqrt{1-y^\gamma}},
\end{equation}
respectively. Thus, it can be seen that for larger values of $\gamma$, $\sigma$ will increase, and ${\cal P}$ and
$\Xi$ will decrease. $\sigma$ is positive for $y<x^{1/\gamma}$ (which is smaller than $x$ since $x<1$ and 
$1/\gamma>1$), and the region of positive $\sigma$ is larger for larger values of $\gamma$. It can be seen in figure \ref{fig5} that $a_0\sigma$ is usually smaller than $a_0{\cal P}$, and it can be negative for values of $x\sim1$. That is because in order to have positive $\sigma$ one needs to restrict attention to $y<x^{1/\gamma}$.
On the other hand, to study the stability of this shell, we replace equations (\ref{bin1}) and (\ref{Phiin1}) in equation
(\ref{master1}) and (\ref{master2}). For the first master equation we get 
\begin{equation}\label{minh1}
 a_0\,m_s''(a_0)\geq\frac{1}{4}\left[\frac{x^2}{(1-x)^{3/2}}-
\gamma\,y^\gamma\frac{(2-\gamma)y^\gamma-2(1-\gamma)}{(1-y^\gamma)^{3/2}}\right],
\end{equation}
for $y<x^{1/\gamma}$. Thus, the stability region is above the clearer surface shown in figure \ref{fig6}. 
In second place, the second master equation is
\begin{equation}\label{minh2}
 a_0^3(4\pi a_0\Xi)''\leq \frac{\gamma}{8}\left[\frac{8}{(1-y^\gamma)^{1/2}}+\frac{6\gamma y^\gamma}{(1-y^\gamma)^{3/2}}
+\frac{\gamma^2 y^\gamma (y^\gamma+2)}{(1-y^\gamma)^{5/2}}\right] ,
\end{equation}
if
\begin{equation}\label{minh2b}
 \frac{\gamma}{2\,r}\frac{2y^\gamma-1}{1-y^\gamma}>0.
\end{equation}
As depicted in figure \ref{fig7}, this quantity is positive if $r_0/a_0$ is not small enough.
Taking into account both stability regions, inequalities (\ref{minh1}) and (\ref{minh2}),
it can be seen that the most stable configurations would have $2M/a_0$ not to close to $1$
(because if not the first stability region would disappear) and intermediate-large
values of $r_0/a_0$ (since this quantity has to be smaller than $2M/a_0$, but not too small to have
a larger second stability region). 
The first (second) stability region would be larger (smaller) for larger values of $\gamma=A\,r_0^2$. The stability region resulting when considering both constraints is
depicted in figure \ref{fig6}.

%%%%%%%%%%%%%%%%%%%%%%%%%%%%%%%%%%%%%%%%%%%%%%%%%%%%%%%%%%%%%%%%%%%%%%%%%%%%%%%%%%%%%%%%%
%%%%%%%%%%%%%%%%%%%%%%%%%%%%%%%%%%%%%%%%%%%%%%%%%%%%%%%%%%%%%%%%%%%%%%%%%%%%%%%%%%%%%%%%%

\subsection{Constant redshift function}\label{inhconstant}

If we consider $\Phi=\Phi_0$ in equation (\ref{binh}), one obtains
\begin{equation}\label{binh1}
 b(r)=\left(1+\frac{A\,r_0^2}{2-\alpha}\right)r-\frac{A\,r_0^\alpha}{2-\alpha}r^{3-\alpha},
\end{equation}
for $\alpha\neq2$, and
\begin{equation}\label{binh2}
 b(r)=r\left[1-A\,r_0^2\,{\rm ln}\left(r/r_0\right)\right],
\end{equation}
for $\alpha=2$. One can easily check that $0<b'(r_0)=1-Ar_0^2<1$, for both cases.
It can be noted that $b(r)$ does not tend to a constant value for $r\rightarrow\infty$.
However, from equation (\ref{binh1}) one can conclude that $b(r)/r$ goes to a constant value in this limit for
$\alpha>2$, whereas it diverges for $0<\alpha<2$. The metric can be expressed as
\begin{equation}\label{monopole2}
 ds^2=-e^{2\Phi_0}dt^2+\frac{\alpha-2}{A\,r_0^2}\frac{dr^2}{1-\left(r_0/r\right)^{\alpha-2}}+r^2 d\Omega_{(2)}^2.
\end{equation}
Therefore, the case $\alpha>2$ is of special interest, as the asymptotic region is particularly simple for this case.
In this case the asymptotic behaviour is that of a global monopole \cite{Barriola:1989hx}, that is,
defining $\widebar r^2=[(\alpha-2)r^2]/(A\,r_0^2)$ and $\widebar t={\rm exp}(\Phi_0)\,t$ in the asymptotic limit we have
\begin{equation}\label{monopole}
 ds^2=-d\widebar{t}^2+d\widebar{r}^2+\frac{A\,r_0^2}{\alpha-2}\widebar{r}^2 d\Omega_{(2)}^2,
\end{equation}
which describes a space with a deficit of solid angle of $\Delta=1-\frac{A\,r_0^2}{\alpha-2}$
for $\alpha>3$ and for $3>\alpha>2$ with $Ar_0^2<\alpha-2$, or an excess of solid angle $-\Delta$ 
otherwise \cite{Barriola:1989hx}. Therefore, metric (\ref{monopole})
may be interpreted as a wormhole carrying a global monopole when $\alpha>2$. 
Anyway, we have obtained an ``almost'' asymptotically Minkowski behaviour in this case. It is interesting to note that previously to Ref. \cite{Barriola:1989hx}, the metric (\ref{monopole}) was studied in \cite{SokStarob} from a purely geometrical point of view, i.e., without relation to a physical monopole.

Taking into account (\ref{binh1}) in equations (\ref{rho})--(\ref{pt}), we get
\begin{equation}
 \rho=\frac{1}{8\pi (\alpha-2)}\left[-A(\alpha-3)\left(\frac{r_0}{r}\right)^\alpha+\frac{\alpha-2}{r^2}-A\left(\frac{r_0}{r}\right)^2\right],
\end{equation}
\begin{equation}
 p_r=\frac{1}{8\pi (\alpha-2)}\left[-A\left(\frac{r_0}{r}\right)^\alpha-\frac{\alpha-2}{r^2}+A\left(\frac{r_0}{r}\right)^2\right],
\end{equation}
\begin{equation}
 p_t=\frac{A}{16\pi}\left(\frac{r_0}{r}\right)^\alpha,
\end{equation}
for $\alpha\neq2$, which decays as $\sim1/r^2$ for $\alpha>2$ as in the case shown in~\cite{Barriola:1989hx}. 
Thus, the transverse pressure is always positive and tends to zero in the asymptotic limit,
as $\rho$ and $p_r$. Moreover, it can be seen that $\rho(r)$ is a decreasing function in the 
interval $(r_0,\,\infty)$ for $\alpha>3$ and for $3>\alpha>2$ with $Ar_0^2<\alpha-2$. Thus,
we have $\rho(r)+p_t(r)\geq0$ which, together with the equation of state (\ref{inh}),
imply that the QNEC is satisfied in the whole space.

On the other hand, for self-completeness, we present the results for the specific case of $\alpha=2$. The metric is given by
\begin{equation}
 ds^2=-e^{2\Phi_0}dt^2+\frac{dr^2}{A\,r_0^2\,{\ln(r/r_0)}}+r^2 d\Omega_{(2)}^2,
\end{equation}
which also leads to $g_{rr}\rightarrow0$ when $r\rightarrow\infty$ as in the previous case for $0<\alpha<2$. 
For this case, the stress-energy tensor profile is given by
\begin{equation}
 \rho=\frac{1-A\,r_0^2}{8\pi r^2}\left[1+{\ln}\left(\frac{r_0}{r}\right)\right], \qquad  p_r=-\frac{1-A\,r_0^2}{8\pi r^2}{ \ln}\left(\frac{r_0}{r}\right), \qquad p_t=-\frac{1-A\,r_0^2}{16\pi r^2},
\end{equation}
respectively.

%%%%%%%%%%%%%%%%%%%%%%%%%%%%%%%%%%%%%%%%%%%%%%%%%%%%%%%%%%%%%%%%%%%%%%%%%%%%%%%%%%%%%%%%%
%%%%%%%%%%%%%%%%%%%%%%%%%%%%%%%%%%%%%%%%%%%%%%%%%%%%%%%%%%%%%%%%%%%%%%%%%%%%%%%%%%%%%%%%%

\section{Discussion and conclusions}\label{summary}

In this work, we have presented new wormhole solutions fuelled by a matter content that minimally violates the null energy condition. 
We have been motivated by a recently proposed cosmological event, denoted ``the little sibling of the big rip'',  where the Hubble rate and the scale factor blow up but the cosmic derivative of the Hubble rate does 
not \cite{Mariam}, as it shows that allowing small bounded violations of the NEC can have relevant
effects in the corresponding geometry.
More specifically, we considered an equation of state in which the sum of the energy density and radial 
pressure is proportional to 
a constant with a value smaller than that of the inverse area characterising the system, i.e., 
the area of the wormhole mouth. Using the cut-and-paste approach, 
we matched interior spherically symmetric wormhole solutions to an exterior Schwarzschild geometry
to obtain asymptotically flat solutions satisfying the QNEC.
We also analysed the stability of the thin-shell to linearized spherically symmetric perturbations around static solutions, 
by choosing suitable properties for the exotic material residing on the junction interface radius. 

Furthermore, we also considered an inhomogeneous generalisation of the equation of state considered  
above. On one hand, we obtain a particular wormhole solution and also match this wormhole geometry 
to an exterior Schwarzschild solution, analysing
the respective stability regions. 
On the other hand, we obtained a specific wormhole solution with an asymptotic behaviour 
corresponding to a space with a deficit of solid angle. This space may, therefore, be interpreted as a
wormhole carrying a global monopole by analogy with \cite{Barriola:1989hx}. This solution
only violates the NEC in a small bounded way through the whole space.

We should refer that in the cases where the wormhole geometries were matched to exterior Schwarzschild solutions, one could construct stable configurations for fluids minimally violating the null energy conditions, 
if the parameters of the model $a_0$, $2M$ and $r_0$ are suitably chosen. In particular, taking into account both master equations, the stability regions would be larger for larger values of $r_0/a_0$. 
However, this quotient should be smaller than $2M/a_0$, which cannot be too close to one. Thus, one could  consider both quotients with large-intermediate values which would lead to sufficiently large stability regions.

On the other hand, it can be noted that the specific equation of state considered in this paper does not allow the existence of static black hole solutions for any sign of $A$. 
This conclusion can be extracted studying the conditions presented in~\cite{Bronnikov:2009ui,Zaslavskii:2010xm} for the existence of these kind of solutions, or simply noting that in order to have a black hole 
in equilibrium with its environment (static solution) one needs a fluid which vanishes at the horizon or behaves as a cosmological constant there, both cases characterized by $p_r+\rho=0$.
Therefore, wormholes seems to be astrophysical objects more natural than black holes in this scenario. Nevertheless, as the existence of black holes is supported by observational data, it would be of
particular interest to study the possible existence of black hole mimickers in this scenario \cite{gr-qc/0012094,gr-qc/0109035,gr-qc/0405111,gr-qc/0407075,quasi1,quasi2,black-star,Visser:2009pw,Lobo:2006xt}.

\begin{acknowledgments}
We thank Matt Visser for helpful comments. The work of MBL was supported by the Basque Foundation for Science IKERBASQUE and the Portuguese Agency ``Funda\c{c}\~{a}o para a Ci\^{e}ncia e Tecnologia" through an Investigador FCT Research contract, with reference IF/01442/2013/CP1196/CT0001. She also wishes to acknowledge the support from the Portuguese Grants PTDC/FIS/111032/2009 and PEst-OE/MAT/UI0212/2014  and the partial support from the Basque government Grant No. IT592-13 (Spain).
 FSNL acknowledges financial  support of the Funda\c{c}\~{a}o para a Ci\^{e}ncia e Tecnologia through an Investigador FCT Research contract, 
with reference IF/00859/2012, funded by FCT/MCTES (Portugal), and grants CERN/FP/123618/2011 and EXPL/FIS-AST/1608/2013. 
PMM also acknowledges financial support of the grant PTDC/FIS/111032/2009 and EXPL/FIS-AST/1608/2013 and thanks 
the hospitality of the University of the Basque Country during the completion of part of this work. 
\end{acknowledgments}

\end{document}